\documentclass[11pt,twoside]{article}
\usepackage{asp2010}
\aspcpryear{2010}
\aspvoltitle{UP: Have Observations Revealed a Variable Upper End of
  the Initial Mass Function?}
\resetcounters

\begin{document}

\title{Applications of the IGIMF-theory  to the astrophysics of galaxies}
\author{Jan Pflamm-Altenburg$^1$, Carsten Weidner$^2$, and Pavel Kroupa$^1$
\affil{$^1$Argelander-Institut f\"ur Astronomie (AIfA), Universit\"at Bonn, 
  Auf dem H\"ugel 71, D-53121 Bonn, Germany}
\affil{$^2$Scottish Universities Physics Alliance (SUPA), School of Physics and
Astronomy, University of St. Andrews, North Haugh, St. Andrews, Fife KY16
9SS, UK}}

\begin{abstract}
The functional form of the galaxy-wide stellar initial mass function
is of fundamental importance for understanding galaxies. 
So far this stellar initial mass function 
has been assumed to be identical to the 
IMF observed directly in star clusters. 
But because stars form predominantly 
in embedded groups rather than uniformly distributed over the whole 
galaxy, the galaxy-wide IMF needs to be calculated by adding all IMFs 
of all embedded groups. This integrated galactic stellar initial mass 
function (IGIMF) is steeper than the canonical IMF and 
steepens with decreasing SFR, leading to fundamental new insights and 
understanding of star forming properties of galaxies.
 This contribution reviews the existing applications of the IGIMF theory to galactic astrophysics, while the parallel contribution by Weidner, Pflamm-Altenburg \& Kroupa (this volume) introduces the IGIMF theory. 
\end{abstract}

\section{Introduction}
Since Salpeters fundamental study it is known that stars do not form
with arbitrary masses, but their masses follow a universal distribution 
function, which is called the initial stellar mass function (IMF).
The IMF, $\xi(m)$, determines the number of newly formed stars, $dN$, 
in the mass interval [$m,m+dm$], and is mathematically defined by 
$\xi(m)=dN/dm$. 

Following its definition, i.e. counting stars, the IMF
has only been determined directly in individual star clusters and 
clusterings.
But the interpretation of galaxy-wide observable quantities, e.g.
chemical abundances or luminosities in different pass bands, 
and the transformation of them into physical quantities such as
total stellar masses, star formation rates, or chemical evolution histories
require the knowledge of the galaxy-wide mass function of all newly formed stars.

Given the lack of galaxy-wide star number counts it has been assumed that 
the galaxy-wide IMF is identical to the IMF in individual star clusters.
But the fact that all stars form in a clustered mode\footnote{Only about 10 per cent of the "embedded clusters" survive to become bound long-lived star clusters.} requires the galaxy-wide IMF
to be calculated by adding all IMFs of all embedded star clusters
 or groups. This leads to 
the concept of the integrated galactic stellar initial mass function (IGIMF).
It is steeper above about 1.3~$M_\odot$
than the individual IMF in the embedded star clusters and steepens with
decreasing SFR. The functional form of the IMF in each
star cluster is universal. But the combination of two facts, i) that only high-mass star
clusters host high-mass stars, and ii) that only high-SFR galaxies host high-mass star
clusters leads to a decrease of the high-mass star fraction and a steepening of the IGIMF
with decreasing SFR (Fig 1, left)\footnote{A tool
  to calculate an IGIMF and to fit it with a multi-part power law as
  required as the input IMF for P{\sc egase} has been presented in
  \citet{pflamm-altenburg2009a} and can be downloaded from
  \tt{www.astro.uni-bonn.de}.}. This decrease of the OB-star fraction with decreasing SFR is called the IGIMF-effect.

\begin{figure}
\plottwo{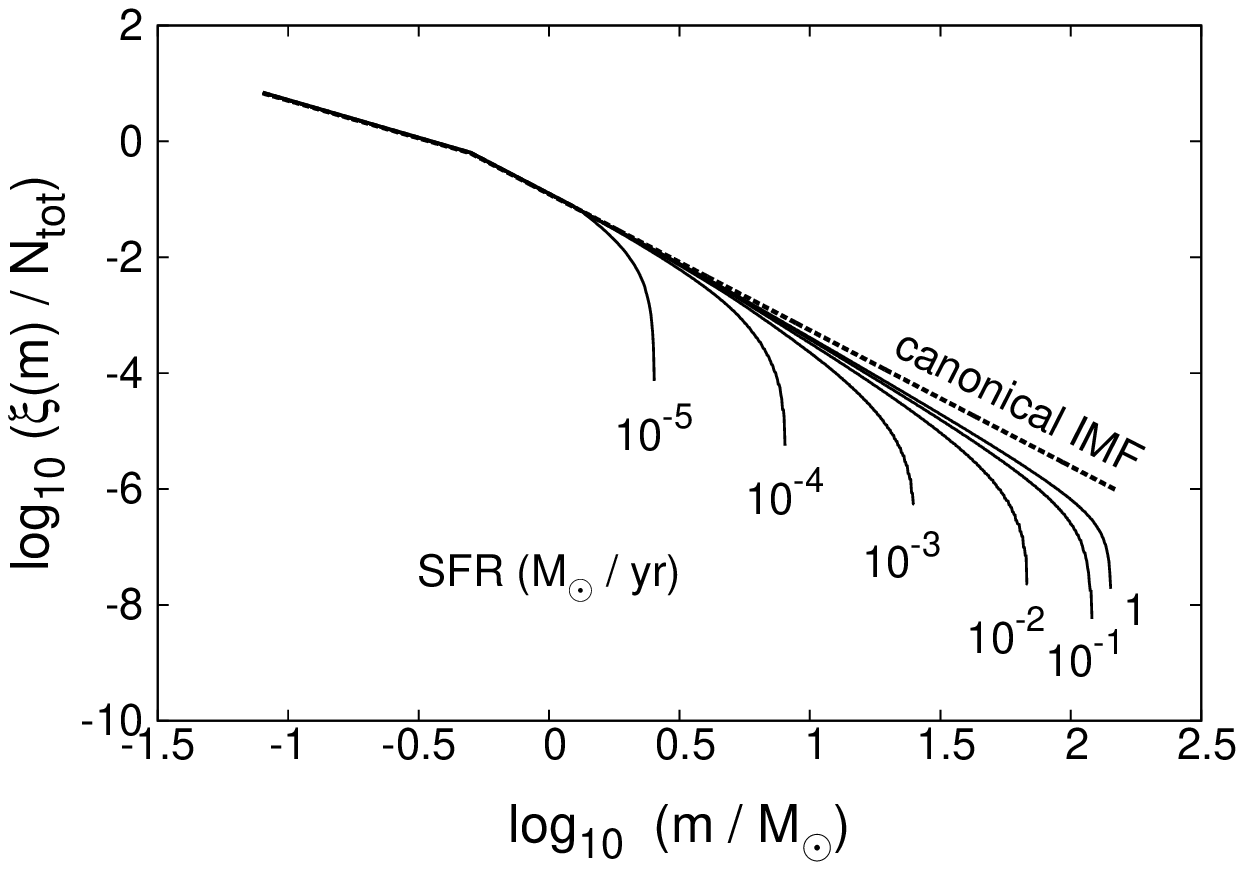}{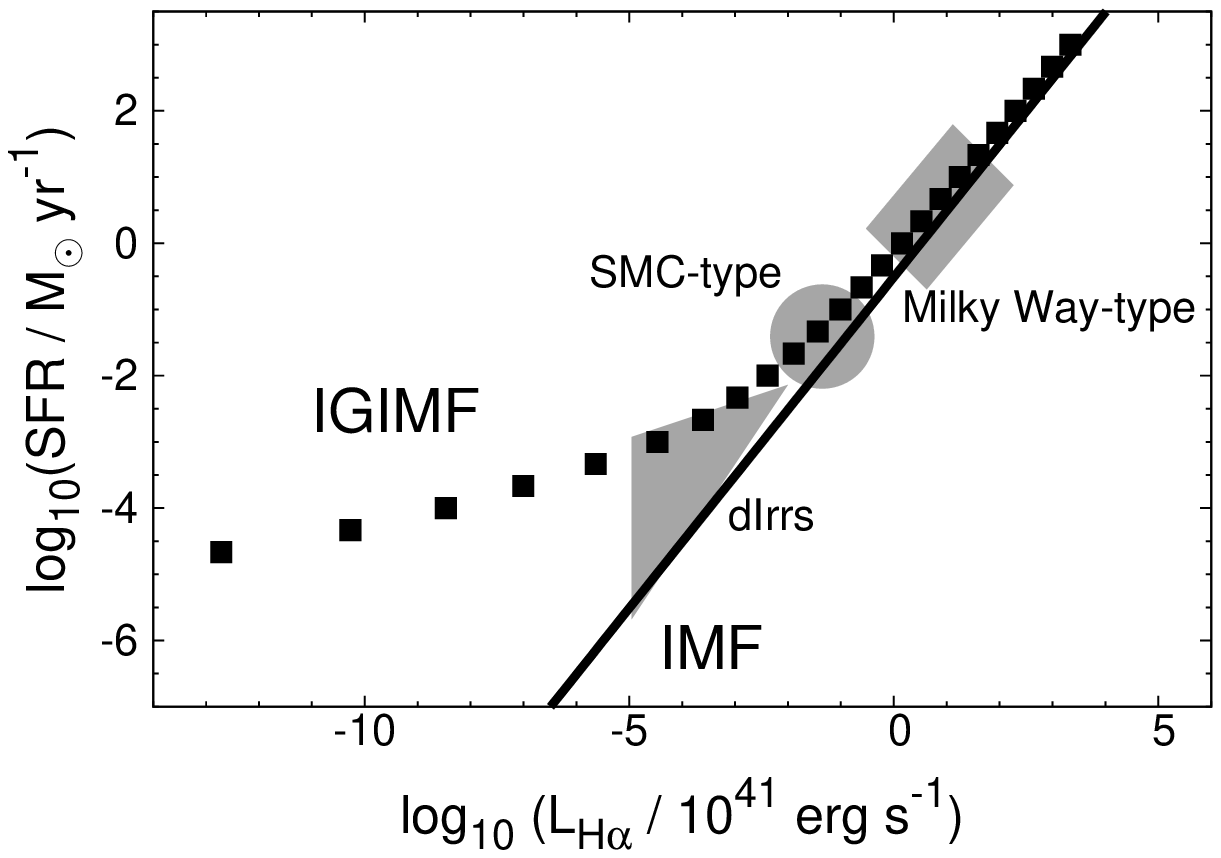}
\caption{\label{fig_igimf} Left: The IGIMF for different SFRs in $M_\odot$/yr. Each IGIMF is normalised such that $\int \xi_\mathrm{IGIMF}(m)\;dm=1$. Right: The recalibrated H$\alpha$-SFR tracer.}
\end{figure}

\section{IGIMF - Applications}
In the following we summarise all applications where the 
constant galaxy-wide IMF has already been replaced by the
IGIMF and list them in Tab.~\ref{tab_igimf_applciations}.

\begin{table}[!ht] 
  \caption{\label{tab_igimf_applciations}IGIMF - applications} 
  \smallskip 
  \begin{center} 
    {\small 
      \begin{tabular}{cc}
        \tableline 
        \noalign{\smallskip} 
        
        Application & Reference\\
        \noalign{\smallskip} 
        \tableline 
        \noalign{\smallskip} 
        $L_\mathrm{H\alpha}$--SFR relation & \citet*{pflamm-altenburg2007d}\\
        
        SFR--$M_\mathrm{gas}$ relation & \citet{pflamm-altenburg2009c}\\
        
        Gas depletion time scales & \citet{pflamm-altenburg2009c}\\
        
        ${H\alpha}$/FUV--SFR relation & \citet*{pflamm-altenburg2009a}\\
        
        Mass--metallicity relation& \citet*{koeppen2007a}\\
        
        [$\alpha$/Fe]--velocity dispersion relation & \citet{recchi2009a}\\

        Colours of star-forming galaxies & \citet{haas2010a}\\
        
        Solar neighbourhood properties & \citet{calura2010a}\\
        Radial H$\alpha$ cutoff in disk galaxies & \citet{pflamm-altenburg2008a}\\
        \noalign{\smallskip} 
        \tableline 
      \end{tabular} 
    } 
  \end{center} 
\end{table}

\subsection{The $L_\mathrm{H\alpha}$--SFR relation}
Because the H$\alpha$ luminosity, $L_\mathrm{H\alpha}$, 
of star forming galaxies depends on the
production rate of ionising photons by high-mass stars, it is expected that 
the usage of $L_\mathrm{H\alpha}$ as a star formation tracer 
needs recalibration in the IGIMF context.
This recalibration is presented in \citet{pflamm-altenburg2007d}
and can be seen in Fig.~\ref{fig_igimf} (right). Milky Way-type galaxies have the 
same H$\alpha$ luminosity for the same SFR in both contexts, constant IMF or 
IGIMF (grey shaded rectangle). The differences start to appear for SMC-type galaxies (grey shaded circle). The  SFRs of dIrrs with the faintest H$\alpha$ 
luminosities are underestimated by up to two orders of magnitude
(gray shaded triangle).

\subsection{The SFR--$M_\mathrm{gas}$ relation of star forming galaxies}
\begin{figure}
\plottwo{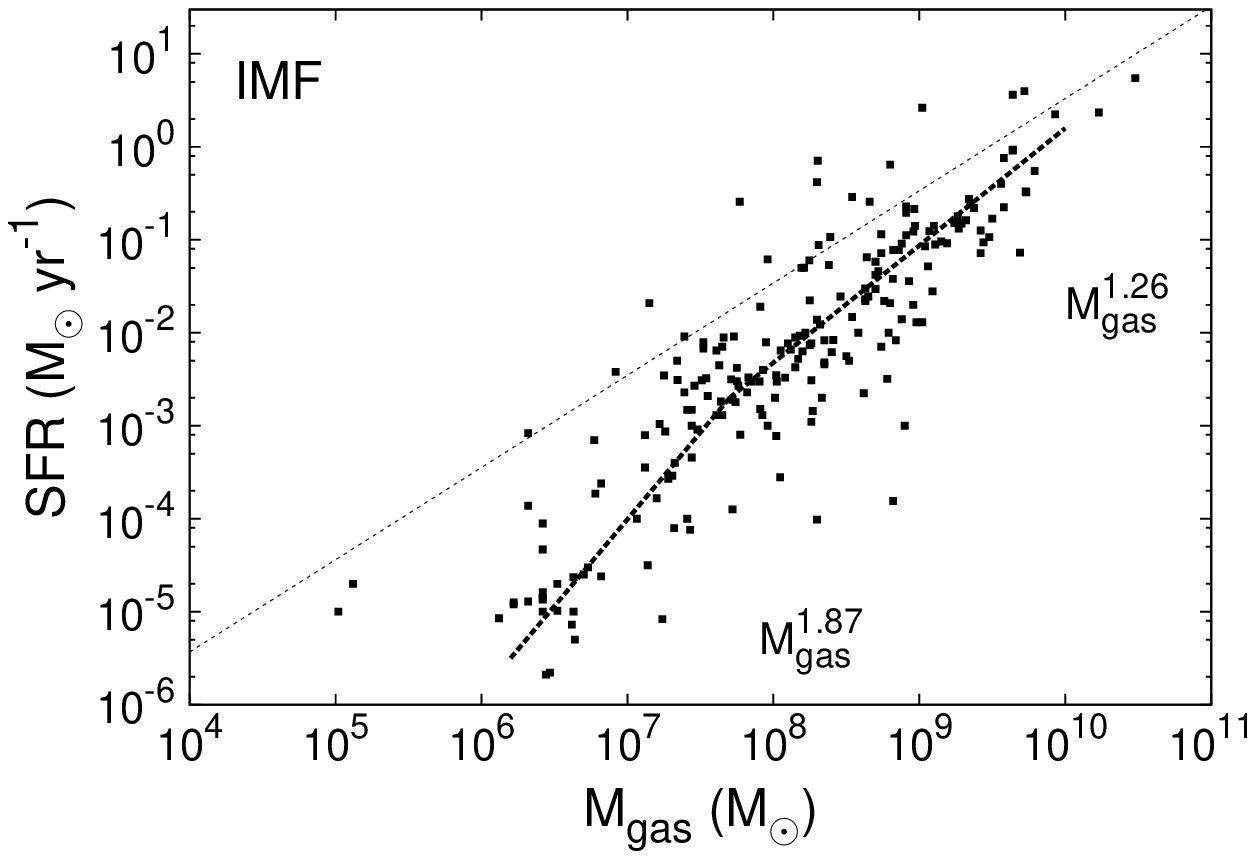}{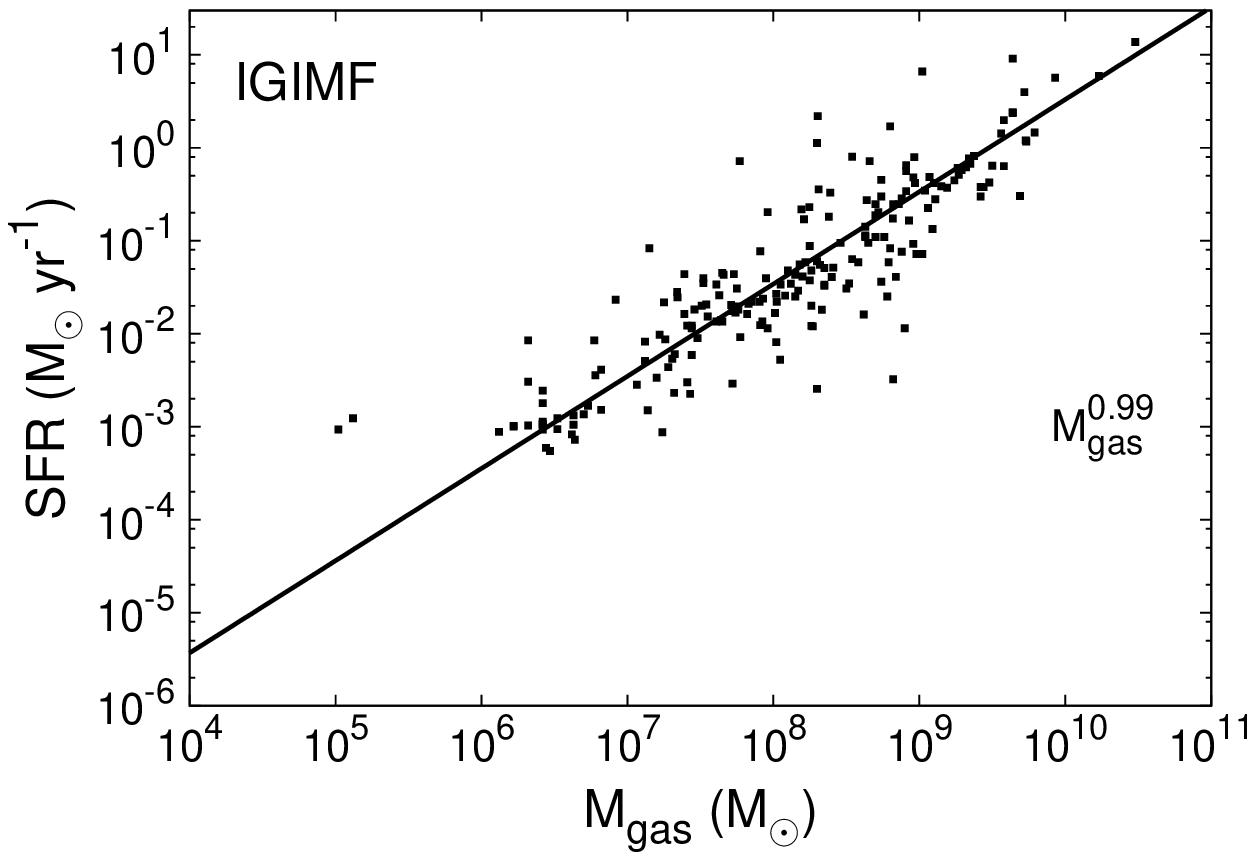}
\caption{\label{fig_sfr} The SFR of galaxies based on H$\alpha$ luminosities
as a function of their total gas mass (from \citealp{pflamm-altenburg2009c}).
Left: For the case of a constant galaxy-wide IMF. Right: For the IGIMF-theory.}
\end{figure}

In the classical constant galaxy-wide IMF picture, total H$\alpha$ luminosities
transform linearly into SFRs. It is found that in this case SFRs 
of galaxies decrease faster than their total gas mass. Furthermore 
the SFR--$M_\mathrm{gas}$ relation steepens  for galaxies
less massive than SMC-type galaxies (Fig.~\ref{fig_sfr}, left). 
In the IGIMF-theory the SFR--$L_\mathrm{H\alpha}$ relation becomes non-linear 
for galaxies which are fainter in H$\alpha$ than SMC-type galaxies.
When using the recalibrated H$\alpha$ star formation tracer in order to correct
for the IGIMF-effect the IGIMF-theory reveals a linear
SFR--$M_\mathrm{gas}$ relation over four orders of magnitude of the gas mass
(Fig.~\ref{fig_sfr}, right).

\subsection{Gas depletion time scales of star forming galaxies}
As a consequence of the revised H$\alpha$ star formation rates in the IGIMF theory, the corresponding gas depletion time scales of star forming galaxies change,
too.
For an assumed constant galaxy-wide IMF the gas depletion time scale, $\tau_\mathrm{gas} = M_\mathrm{gas}$/SFR, increases with decreasing galaxy gas mass 
as a result of a steeper than linear SFR--$M_\mathrm{gas}$ relation.
This has lead to the interpretation that dwarf irregular galaxies are inefficient
in forming stars compared to large disk galaxies.
The IGIMF based SFR--$M_\mathrm{gas}$ relation is linear and therefore the 
gas depletion time scales are constant at about 3~Gyr \citep{pflamm-altenburg2009c}. In the IGIMF-context all star forming galaxies have the same
star formation efficiency.

\subsection{Stellar-mass buildup times  and downsizing}
The stellar-mass buildup time, $\tau_\mathrm{*}=M_\mathrm{*}/SFR$, is a measure
for the ratio of the average past and the current SFR of a galaxy.
Using the canonical SFRs calculated with an invariant IMF it follows that $\tau_*$ becomes longer than a Hubble time for dwarf galaxies. This is unphysical and is also in contradiction to the observationally found downsizing result. Using instead the IGIMF-based SFRs, $\tau_*$ becomes shorter than a Hubble time for virtually all galaxies and decreases towards less-massive galaxies being consistent with downsizing \citep{pflamm-altenburg2009c}.

\subsection{The ${H\alpha}$/FUV--SFR relation}
\begin{figure}
  \plottwo{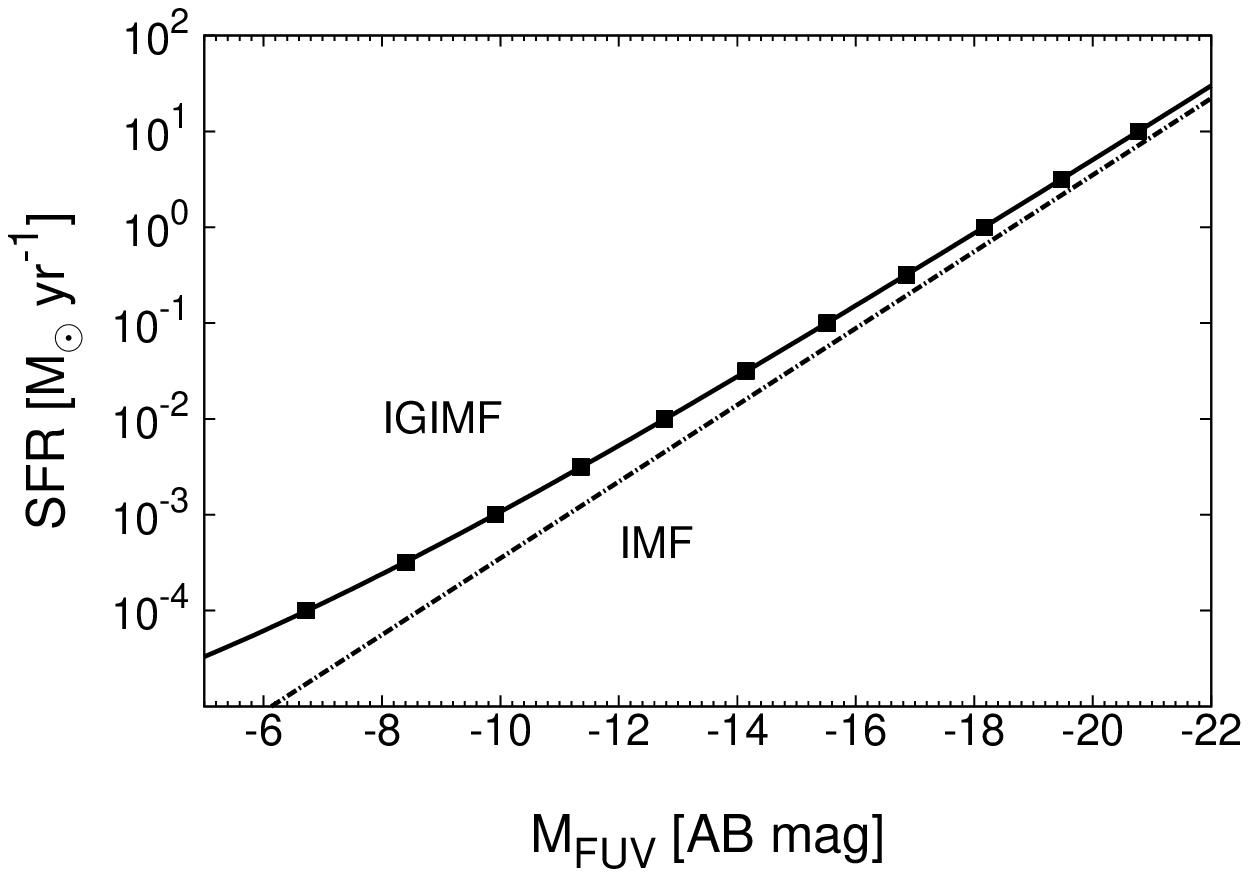}{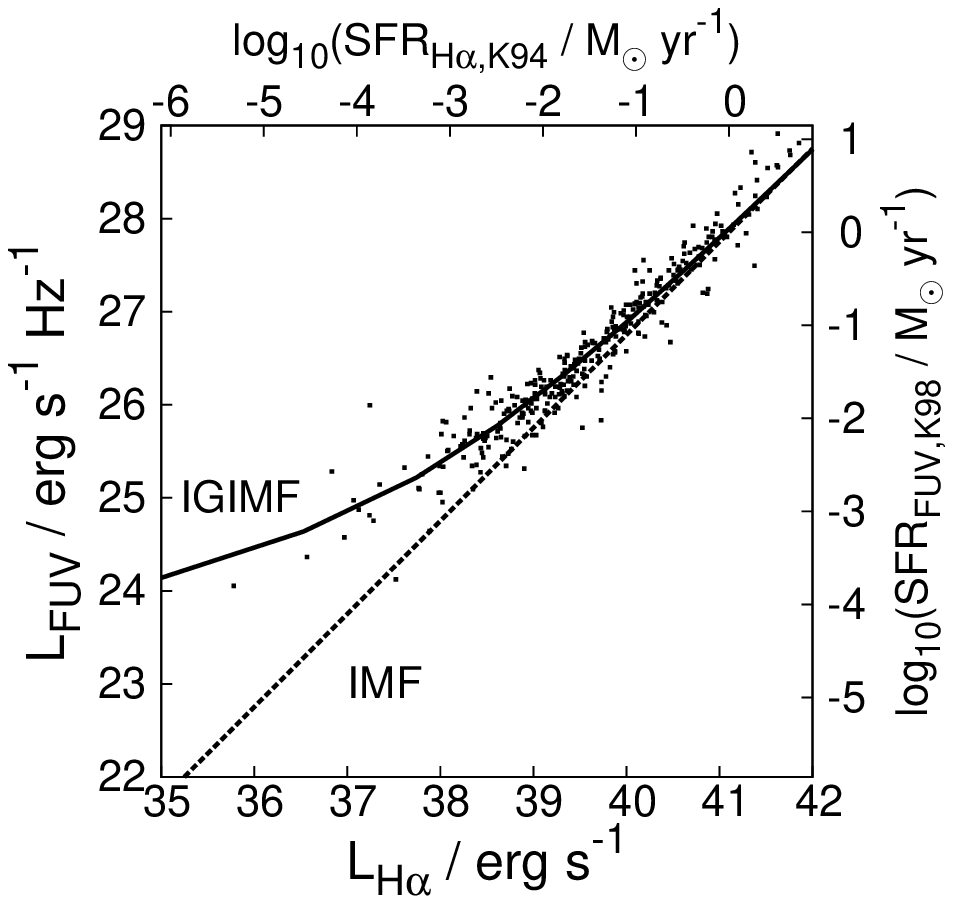}
  \caption{\label{fig_fuv} Left: The SFR--FUV-magnitude relation for the constant IMF case and predicted by the IGIMF-theory for solar metallicity
\citep{pflamm-altenburg2009a}.
Right: H$\alpha$ and FUV luminosities expected in the case of a constant IMF 
and predicted by the IGIMF-theory
compared with observation by \citet{lee2009a} of local volume star-forming 
galaxies.}
\end{figure}

The H$\alpha$ luminosity is produced by recombining hydrogen ionised by the
Lyman $\alpha$ continuum photons of high-mass stars and shows therefore a
strong IGIMF-effect. Contrary, the far ultraviolet (FUV) luminosity of galaxies 
is produced mainly by long-lived B-stars. Thus, the FUV-flux of galaxies
also shows an IGIMF-effect (Fig.~\ref{fig_fuv}, left) but less so than 
the H$\alpha$ luminosity. E.g. for a default SFR of 
10$^{-3}$~M$_\odot$~yr$^{-1}$ (10$^{-4}$~M$_\odot$~yr$^{-1}$) the IGIMF-effect 
for H$\alpha$ is 2.0~dex (5.0~dex), whereas the IGIMF-effect for
FUV is 1.22~mag (1.91~mag) or 0.48~dex (0.77~dex) (compare Fig.~\ref{fig_igimf} right and Fig.~\ref{fig_fuv} left). In this context the IGIMF-effect means the difference of the logarithm of the H$\alpha$ luminosity and FUV-luminosity 
between the IGIMF model and the constant-IMF case.

It should be noted that due to the IGIMF-effect in the FUV-luminosity the
SFRs of galaxies which are calculated from FUV-luminosities and a classical
linear FUV-luminosity--SFR relation resulting from an assumed constant IMF
are closer to the true underlying SFR than H$\alpha$ based SFRs,
but they are still lower.

Combining the $L_\mathrm{FUV}$--SFR and the $L_\mathrm{H\alpha}$--SFR relations
predicted by the IGIMF-theory a relation between the 
$L_\mathrm{H\alpha}$/$L_\mathrm{FUV}$-ratio and the H$\alpha$ luminosity
has been predicted
in \citet{pflamm-altenburg2009a}. This prediction is in quantitative 
agreement with the observations by \citet{lee2009a} (Fig.~\ref{fig_fuv}, 
right).

\subsection{The mass--metallicity relation}
\begin{figure}
  \plottwo{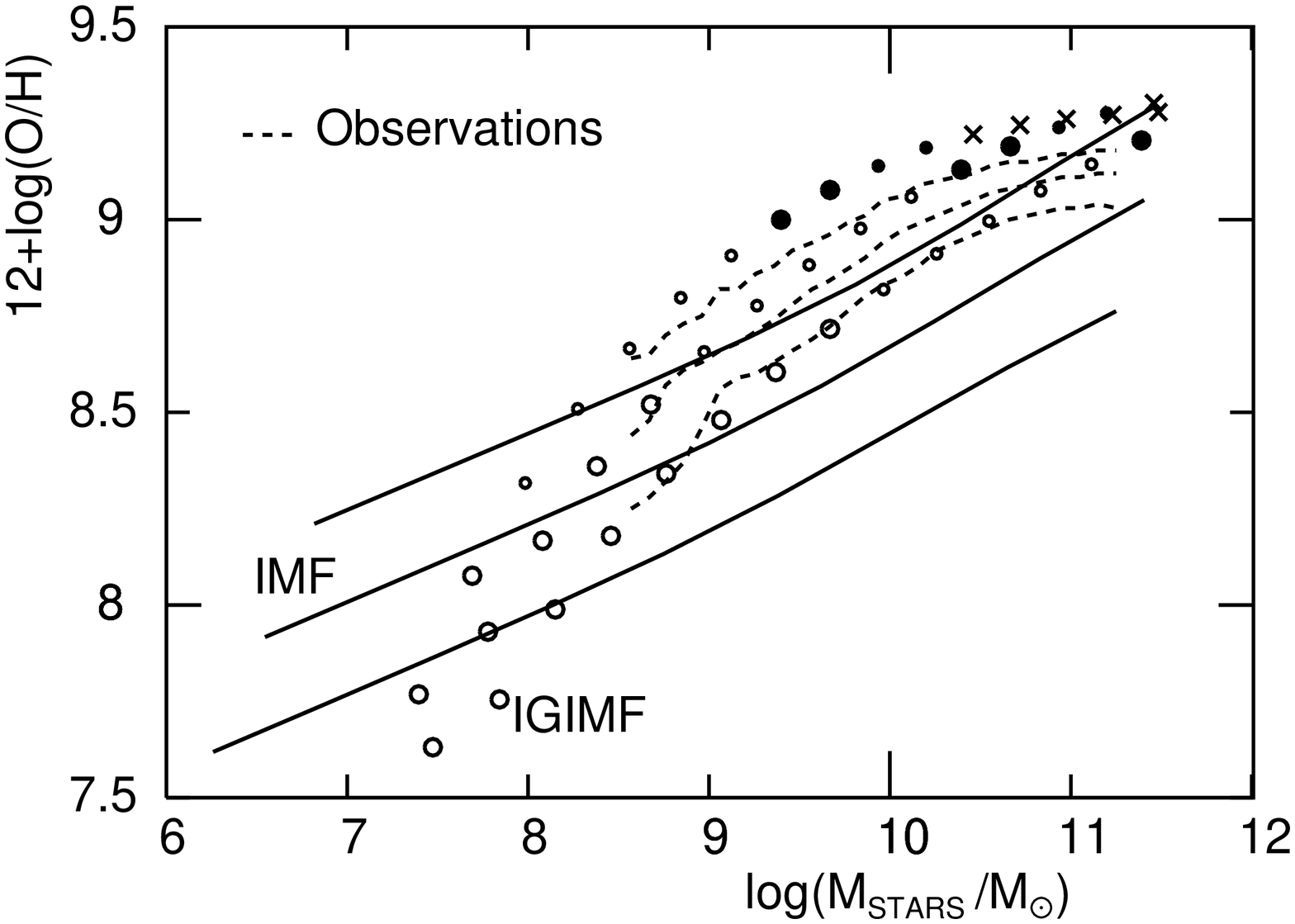}{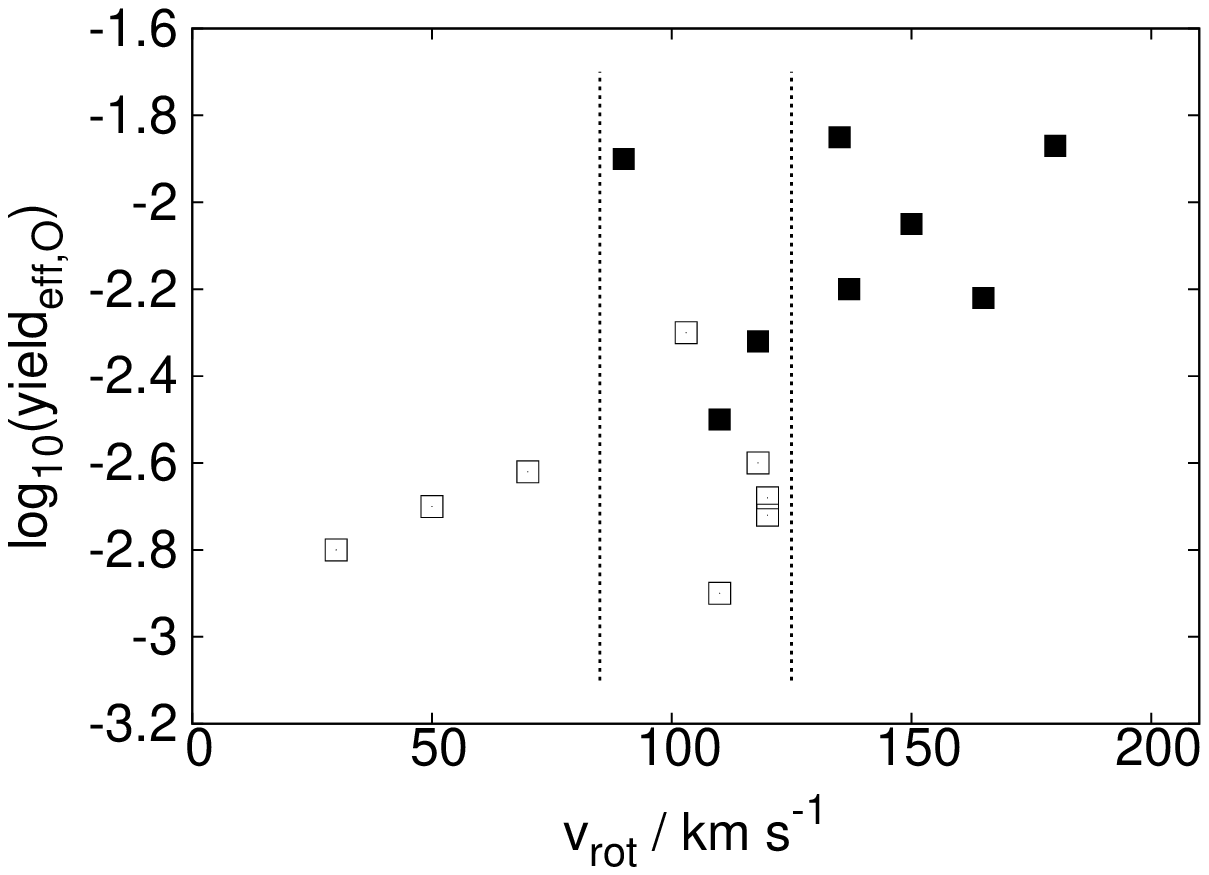}
  \caption{\label{fig_metal} Left: The mass--metallicity relation of galaxies in the IGIMF-theory (symbols) as calculated in \citet{koeppen2007a} 
compared with the
observations by \citet{tremonti2004a} (dashed lines). 
Right: Effective oxygen yields of LSBs (open squares)
and normal spiral galaxies (black squares) based on a data compilation from \citet{garnett2002a}.}
\end{figure}

The IGIMF-effect describes the decreasing high-mass star fraction of all newly formed stars with decreasing SFR. Thus the amount of freshly produced chemical 
elements, which are synthesised by high-mass stars (e.g. oxygen in SNII),
per total newly formed stellar mass must decrease with decreasing
SFR. It is thus expected within the IGIMF theory 
that the metallicity as well as the detectable, i.e. effective, yield of oxygen
of galaxies increase with increasing galactic SFR and therefore stellar mass. 
\citet{koeppen2007a} have shown that both relations
are a natural consequence of the IGIMF-theory and agree with the observations
(Fig~\ref{fig_metal}, left).
In order to explain the observed mass-metallicity relation of galaxies
with a constant galaxy-wide IMF the occurrence of 
metal-enriched outflows has  to be speculated.
This requires that a large fraction of freshly produced metals 
must escape from the galaxy and is in principle equivalent to reducing the
number fraction of high-mass stars as pointed out by \citet{koeppen2007a}. 

On first sight this might be plausible as lower-mass galaxies have shallower
gravitational potentials than large disk galaxies, and expanding supernova shells
containing freshly produced metals might escape easier from dwarf galaxies.
In order to break the degeneracy between the IGIMF-model, on the one-hand side, 
 and the constant 
IMF model plus metal enriched outflows, on the other hand side, 
 one has to concentrate on galaxies 
with different SFRs but equal gravitational potentials. 
This can be done by
comparing low-surface brightness galaxies (LSBs) with normal disk galaxies 
having the same rotational velocity which is a proxy for the deepness of the 
gravitational potential. 

For the same potential, the constant-IMF model combined with metal enriched outflows would predict that
the effective yields are higher for LSBs than for normal disk galaxies, because
normal disk galaxies have higher SFRs and larger feedback by supernovae 
and metals should escape easier  reducing the effective, i.e. detectable, 
yields. 
The IGIMF-model, on the other hand,  predicts that the effective
yields are higher for normal disk galaxies than
for LSBs because normal disk galaxies have higher SFRs and therefore flatter
IGIMFs and a larger fraction of high-mass stars.

Fig.~\ref{fig_metal} (right) shows the effective oxygen yields of normal disk galaxies and 
LSBs. At a rotational velocity of $\approx$100~km/s the regimes of normal disk galaxies and LSBs overlap. In the region of the same rotational velocity 
normal disk galaxies have higher effective yields than LSBs in agreement with the IGIMF-theory
(Pflamm-Altenburg \& Kroupa -- in prep.).
\subsection{The [$\alpha$/Fe]--velocity-dispersion relation}
If different chemical elements are produced in stars with different 
masses then the corresponding effective yields must have different IGIMF-effects.
E.g. oxygen (or any $\alpha$-element as e.g Mg) 
is produced in SNII which have high-mass stars
as progenitors, whereas iron is mainly produced in SNIa which 
have white dwarfs as their progenitors. Thus, the oxygen production 
in galaxies should decrease faster with decreasing SFR than the iron production.
In other words the IGIMF-effect for oxygen is stronger than the IGIMF-effect 
for iron. 

In early-type galaxies the observed velocity dispersion scales with the
total stellar mass of the galaxies. As high-mass galaxies must have had a 
higher SFR than low-mass galaxies it is expected in the IGIMF-theory
that the [$\alpha$/Fe] abundance ratio decreases with decreasing 
velocity dispersion of the galaxies. \citet{recchi2009a} have shown 
that the chemical evolution of early type galaxies in the context of 
the IGIMF-theory is in quantitative agreement with the observations.
Furthermore, the IGIMF-models produce a steeper [$\alpha$/Fe]--$\sigma$
relation in low-mass galaxies, as is observed, whereas standard
constant IMF models only 
provide theoretical [$\alpha$/Fe] ratios in agreement with observed 
ratios in high-mass early-type galaxies, only reproduce the
steepening of the [$\alpha$/Fe]--$\sigma$ relation for low-mass 
early-type galaxies  if element-selective outflows are speculated to occur.

\subsection{Solar neighbourhood properties}
The solar neighbourhood is the ideal test place to compare IGIMF-predictions
with observations as it is i) a result of the composition of numerous dissolved
star clusters and thus represents a composite stellar population
in terms of the IGIMF-theory, and ii) due to its proximity
star counts can be performed and metallicity determinations are most accurate.
\citet{calura2010a} vary systematically 
the slope of the embedded cluster mass function, $\beta$,  and calculate
theoretically 
the present-day stellar mass function (PDMF) and chemical enrichment
in the IGIMF-context. A choice of $\beta=2$ leads to the best agreement
between the IGIMF-model predictions and the observed PDMF and 
chemical properties of the solar neighbourhood simultaneously.

\subsection{The colours of starforming galaxies}
The IGIMF-effect of a galaxy-wide integrated property becomes smaller
if the fraction of contributing long-lived low-mass stars increases.
Therefore, optical bass bands should exhibit an IGIMF-effect, too, but less
strongly than  the IGIMF-effect for FUV. First studies to explore the IGIMF-effect 
on widely-used pass bands as e.g. U, B, V have been presented in 
\citet{haas2010a}.
Their main result is that optical colours vary only slightly within different
IGIMF-models and that this variation is smaller than the intrinsic
scatter within a morphological class of galaxies. Therefore, optical colours of
galaxies can not be used to constrain details of the IGIMF theory.

Long-lived low-mass stars are the main contributor to optical luminosities
and therefore the star-formation history (SFH) will have an important influence
on the IGIMF-effect integrated over time. Further studies are required in order
to explore the SFH dependence on galactic colours in the IGIMF context. 

\subsection{The radial H$\alpha$ cutoff in star-forming disk galaxies}
The IGIMF-effect in galaxies is a result of a decreasing mass of the
heaviest star cluster with decreasing SFR. This can be physically
understood as high-SFR galaxies have higher gas densities and more material
is locally available to form high-mass star clusters. 
Within galaxies the gas density decreases in general with increasing
galactocentric radius. Therefore we would expect a \emph{local} IGIMF-effect, 
i.e. a composite IMF of many star forming regions should be steeper in the
outer disk of star forming galaxies than in their inner regions.

A local integrated galactic stellar initial mass function (LIGIMF) can be 
constructed straightforwardly be replacing all relevant quantities in 
the IGIMF formulation by their corresponding surface densities
\citep{pflamm-altenburg2008a}:
The LIGIMF then defines the number of newly formed stars, $dN$, 
in the mass interval from $m$ to $m+dm$ per unit area at the location (x, y) 
in a star-forming disk galaxy and is calculated by adding all IMFs 
of all locally newly formed star clusters
\begin{equation}
  \xi_\mathrm{LIGIMF}(m,x,y)=
  \int_{M_\mathrm{ecl,min}}^{M_\mathrm{ecl,max,loc}(x,y)}
  \xi_\mathrm{M_\mathrm{ecl}}(m)\;\xi_\mathrm{LECMF}\;(M_\mathrm{ecl},x,y)\;
  \mathrm{d}M_\mathrm{ecl}\;,
\end{equation}
where $\xi_\mathrm{LECMF}\;(M_\mathrm{ecl},x,y)$ is the local embedded
cluster mass function surface density (LECMF), 
which defines the number of newly formed star clusters, $dNecl$, 
in the mass interval from $M_\mathrm{ecl}$ to $M_\mathrm{ecl}+dM_\mathrm{ecl}$ 
per unit area at the location (x, y) in a star-forming disk galaxy, and 
$\xi_\mathrm{M_\mathrm{ecl}}(m)$ is the IMF of an embedded star cluster with 
total stellar $M_\mathrm{ecl}$.

In order to express the dependence of the local upper limit of the 
LECMF on the local gas surface density, the ansatz,
\begin{equation}
  M_\mathrm{ecl,max,loc}(x,y) = 
  M_\mathrm{ecl,max}
  \left(\frac{\Sigma_\mathrm{gas}(x, y)}{\Sigma_\mathrm{gas,0}}\right)^\gamma\;,
\end{equation}
is made, where $\Sigma_\mathrm{gas}(x, y)$ and $\Sigma_\mathrm{gas,0}$ are the gas surface
densities at the location (x, y) and at the center of the galaxy and 
$M_\mathrm{ecl,max}$ is the most massive star cluster in the galaxy determined by the total SFR.  

The LECMF is locally normalised such that
\begin{equation}
\delta t\;\Sigma_\mathrm{SFR}(x,y)=
\int_{M_\mathrm{ecl,min}}^{M_\mathrm{ecl,max,loc}(x,y)}\xi_\mathrm{LECMF}{(M_\mathrm{ecl})}\;M_\mathrm{ecl}\;\mathrm{d}M_\mathrm{ecl}\;,
\end{equation}
where $\Sigma_\mathrm{SFR}$ is the star-formation surface density, 
$\delta t\approx10\;\mathrm{Myr}$ is the time span required to
populate the cluster mass function completely \citep{weidner2004b}.

For an exponential gas disk and a Kennicutt-Schmidt law with general exponent $N$ the average radial star-formation rate surface density can be calculated
(see supplement of \citealt{pflamm-altenburg2008a} for details).
As the the gas surface density decreases with increasing galactocentric radius
the local upper mass limit of the LECMF decreases and the local fraction
of high-mass stars decreases, too. Consequently, the LIGIMF
steepens with increasing galactocentric radius and the radial H$\alpha$ 
surface density decreases faster than the radial FUV surface density.

\begin{figure}
  \plottwo{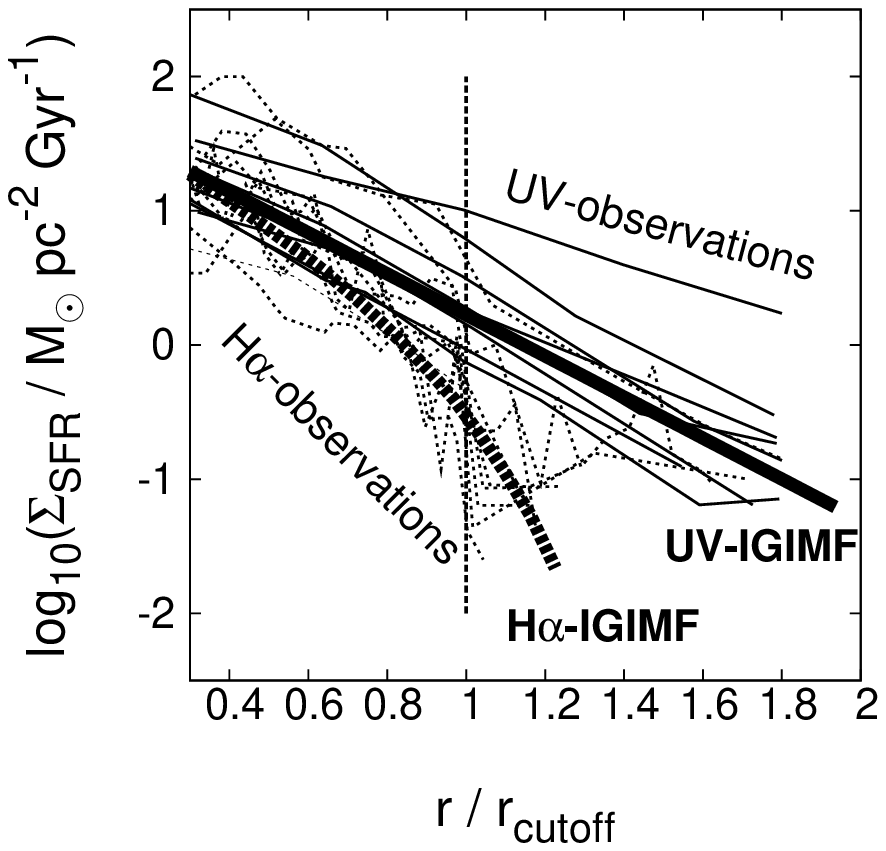}{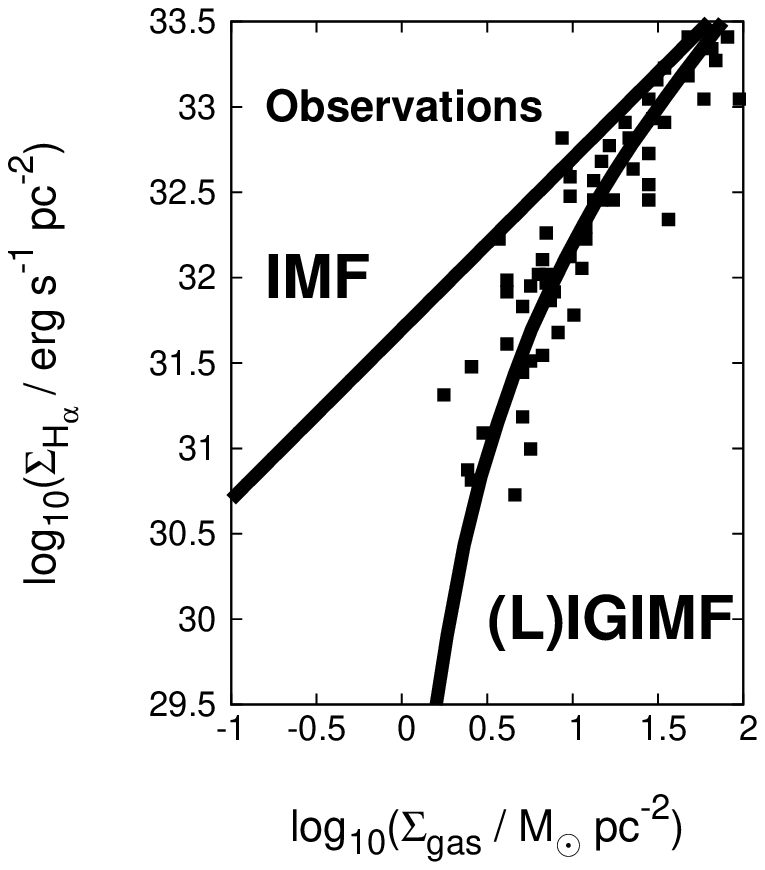}
  \caption{\label{fig_ha_cutoff} Left: Observed \emph{classical} (constant IMF) FUV (thin solid lines) and H$\alpha$ SFRs (thin dotted lines) 
\citep{boissier2007a} 
and expected classical SFRs in the IGIMF-context (thick solid and dotted lines) \citep[from][]{pflamm-altenburg2008a}.
Right: Local observed H$\alpha$ surface densities as function of the local gas surface density (black squares) compared with the expectations from an assumed constant IMF and the IGIMF for a linear Kennicutt-Schmidt law ($N=1$) and no star-formation cutoff \citep[from][]{pflamm-altenburg2008a}.}
\end{figure}

For a choice of $N=1$ and $\gamma=\frac{3}{2}$ of star-forming disk galaxies
an agreement between i) the observed radial H$\alpha$ and FUV-luminosity 
profiles (Fig.~\ref{fig_ha_cutoff} left), and ii) a match of an apparent star
formation cutoff in low gas density environments (Fig.~\ref{fig_ha_cutoff} left)
can be achieved.

\section{Conclusion}
We have presented a summary of astrophysical topics where the constant
galaxy-wide IMF used so far has been replaced by the IGIMF. The common result 
is that the usage of a constant galaxy-wide IMF requires including additional  
assumptions in order to explain or understand physical properties of galaxies.
The IGIMF-theory instead provides natural explanations of galactic 
properties with no need of further ad-hoc assumptions. 
It should be noted that how the IGIMF varies with 
SFR is not arbitrary but rests entirely on the clustered nature  
of star formation deduced from nearby star-forming events 
and is quantitatively determined by the galaxy-wide SFR  or the local SFR density within a galaxy.

\bibliography{pflamm-altenburg_j}

\end{document}